\documentclass[aps,preprint,showpacs,amsmath,amssymb,amsfonts]{revtex4}
\usepackage{epsfig}
\usepackage{graphicx}
\usepackage{subfigure}
\usepackage{dcolumn}
\usepackage{bm}
\usepackage{amsthm}
\usepackage{enumitem}
\usepackage{fixltx2e}
\usepackage{slashed}

\newcommand{\const}{\mbox{const}}
\newcommand{\del}{\partial}

\newcommand{\bbR}{\mathbb{R}}

\begin{document}

\title{Higher-spin gravity as a theory on a fixed (anti) de Sitter background}

\author{Yasha Neiman}
\email{yashula@gmail.com}

\affiliation{Perimeter Institute for Theoretical Physics, 31 Caroline Street N, Waterloo, ON, N2L 2Y5, Canada}

\date{\today}

\begin{abstract}
We study Vasiliev's higher-spin gravity in 3+1d. We formulate the theory in the so-called compensator formalism, where the local isometry group $SO(4,1)$ is reduced to the Lorentz group $SO(3,1)$ by a choice of spacelike direction in an internal 4+1d space. We present a consistent extension of Vasiliev's equations that allows this internal direction to become spacetime-dependent. This allows a new point of view on the theory, where spacetime is identified with the de Sitter space of possible internal directions. We thus obtain an interacting theory of higher-spin massless gauge fields on a fixed, maximally symmetric background spacetime. We expect implications for the physical interpretation of higher-spin gravity, for the search for a Lagrangian formulation and/or quantization, as well as for higher-spin holography.
\end{abstract}


\maketitle
\newpage

\section{Introduction} \label{sec:intro}

Vasiliev's higher-spin gravity \cite{Vasiliev:1995dn,Vasiliev:1999ba} is an interacting theory of massless gauge fields, which include a graviton with spin $s=2$ as well as an infinite tower of fields with increasing spin $s>2$. The theory is known non-perturbatively at the level of classical equations of motion. The equations are invariant under diffeomorphisms, as well as under an infinite-dimensional higher-spin gauge group. It appears that apart from a small set of choices, the higher-spin symmetry uniquely determines the field equations, to all orders in the interaction strength and in spacetime derivatives. This suggests that if the theory can be quantized without breaking its gauge symmetry, the quantization will automatically be free of ambiguities. 

An additional source of interest in higher-spin gravity is that like string theory, it appears to participate in an AdS/CFT duality \cite{Klebanov:2002ja,Sezgin:2003pt,Giombi:2012ms}, with a vector model as the CFT dual. Most importantly, \emph{unlike} string theory, higher-spin gravity can be formulated just as easily with a positive cosmological constant. This allows the construction of a concrete holographic duality in four-dimensional de Sitter space \cite{Anninos:2011ui}. 

At the present state of development, higher-spin gravity is not a realistic model for our Universe: its interactions appear to be non-local at the cosmological scale, and there is no known limit in which they become the local interactions of General Relativity. Such a limit may exist as a broken-symmetry phase, but no concrete mechanism is known so far \footnote{As a counterpoint, note the conjectured duality \cite{Chang:2012kt} between supersymmetric higher-spin gravity and string theory in $AdS_4$. However, this example does not quite count. Indeed, the strings and ultimately General Relativity emerge in \cite{Chang:2012kt} after quantizing the higher-spin theory, and the quantization in the relevant setup intrinsically breaks the higher-spin symmetries. Thus, General Relativity is not obtained as a limit of a theory with higher-spin symmetry.}. Thus, our main motivation for studying this theory is as a model for holography and quantum gravity in the physical spacetime dimension (3+1d) with the physical sign of the cosmological constant ($\Lambda>0$).

In this paper, we present an extension of Vasiliev's higher-spin field equations, in which the so-called compensator vector is allowed to be non-constant in spacetime. The compensator is a non-dynamical unit spacelike vector in an internal 4+1d Minkowski space. Its purpose is to break the local $SO(4,1)$ group of de Sitter isometries down to the Lorentz group $SO(3,1)$. In the existing formulations of higher-spin theory, it is chosen (sometimes implicitly) to be constant. At the linearized level, the equations of motion with a non-constant compensator are known \cite{Vasiliev:2001wa,Didenko:2012vh}, and describe free massless gauge fields as expected. What we accomplish is a marriage between this latter form of the free equations and the full machinery of the non-linear theory. 

The main upshot of our result is that once the compensator becomes spacetime-dependent, its value can be used as a label for the spacetime points themselves. Spacetime is thus identified with the space of possible values of the compensator, which is just pure de Sitter space $dS_4$. In this identification, we sacrifice diffeomorphism invariance, reducing it to the de Sitter isometry group. However, the internal higher-spin gauge symmetry remains intact. Thus, we end up with a non-perturbative interacting theory of massless gauge fields on a fixed Sitter background. In particular, the dynamical fields have a spin-2 component, which simply lives on the de Sitter metric, and cannot be viewed as a perturbation of it.

The picture sketched above is a diametric opposite of the standard formulation of higher-spin gravity. There, spacetime essentially disappears, more so than in General Relativity. Indeed, the metric in the standard picture is merely a component of the higher-spin gauge connection, mixed with other components under gauge transformations. In particular, any notion of asymptotics or horizons in spacetime entails a gauge choice. In addition, Vasiliev's field equations impose flatness on the higher-spin gauge connection, and vanishing covariant derivatives for the other master fields. Fields at different spacetime points are thus related by a gauge transformation, essentially demoting spacetime into a 4d set of gauge frames. In contrast, in our picture, the spacetime metric is fixed, while the higher-spin connection and some of the covariant derivatives become non-trivial.

The paper is organized as follows. In section \ref{sec:spinors}, we outline a formalism for spinors and twistors in global $dS_4$, presented originally in \cite{Neiman:2013hca}. In section \ref{sec:free}, we review the linearized higher-spin equations with and without a spacetime-dependent compensator. In the existing 
treatments \cite{Vasiliev:2001wa,Didenko:2012vh}, a spacetime-dependent compensator obscures the gauge invariance of the equations; we present an alternative formulation that makes the invariance manifest. Using this formulation as a starting point, we construct in section \ref{sec:interacting} the full non-linear equations with a spacetime-dependent compensator. With particular choices for the compensator field, these reduce to Vasiliev's standard formulation or to the formulation on pure de Sitter space discussed above. Section \ref{sec:discuss} is devoted to discussion and outlook. We speculate on the equivalence between our version of higher-spin theory and the standard one, discuss potential implications and list open questions. The consistency of the new non-linear equations is analyzed in detail in the Appendix.

For concreteness and physical relevance, we assume throughout that spacetime is Lorentzian with $\Lambda>0$. However, it should be possible to adapt our results to any spacetime signature and sign of $\Lambda$.

\section{Spinors and twistors in de Sitter space} \label{sec:spinors}

We define de Sitter space $dS_4$ as the hyperboloid of unit spacelike vectors in 4+1d flat spacetime:
\begin{align}
 dS_4 = \left\{v^I\in\bbR^{1,4}\, |\ v_I v^I = 1\right\} \ , \label{eq:dS}
\end{align}
where we chose units so that the cosmological constant is $\Lambda = 3$. The indices $(I,J,\dots)$ take values from $0$ to $4$, and are raised and lowered with the Minkowski metric $\eta_{IJ}$ of signature $(-,+,+,+,+)$. We will use the same indices for vectors $\xi^I$ in the tangent bundle of $dS_4$, with the understanding that these must be tangent to the hyperboloid \eqref{eq:dS}, i.e. $v_I \xi^I = 0$. The de Sitter isometry group $SO(4,1)$ is realized as the rotation group in $\bbR^{1,4}$.

Twistors \cite{Penrose:1986ca,Ward:1990vs} in de Sitter space are the 4-component Dirac spinors of the isometry group $SO(4,1)$. We use $(a,b,\dots)$ for twistor indices. The twistor space has a symplectic metric $I_{ab}$, which we use to raise and lower indices via $Z_a = I_{ab}Z^b$ and $Z^a = Z_b I^{ba}$, where $I_{ac}I^{bc} = \delta_a^b$. Twistor indices and tensor indices in $\bbR^{1,4}$ are related through the gamma matrices $(\gamma_I)^a{}_b$, which satisfy the Clifford algebra $\{\gamma_I,\gamma_J\} = -2\eta_{IJ}$. These 4+1d gamma matrices can be realized as the usual 3+1d ones, with the addition of $\gamma_5$ (in our notation, $\gamma_4$) for the fifth direction in $\bbR^{1,4}$. Concretely, these matrices can be represented in $2\times 2$ block notation as:
\begin{align}
 \begin{split}
   I_{ab} &= -i\begin{pmatrix} 0 & \sigma_2 \\ \sigma_2 & 0 \end{pmatrix} \ ; \\
   (\gamma^0)^a{}_b &= \begin{pmatrix} 0 & 1 \\ 1 & 0 \end{pmatrix} \ ; \quad 
   (\gamma^k)^a{}_b = -i\begin{pmatrix} \sigma^k & 0 \\ 0 & -\sigma^k \end{pmatrix} \ ; \quad 
   (\gamma^4)^a{}_b = \begin{pmatrix} 0 & -1 \\ 1 & 0 \end{pmatrix} \ ,
 \end{split}
\end{align}
where $\sigma_k$ with $k=1,2,3$ are the Pauli matrices. The $\gamma^I_{ab}$ are antisymmetric and traceless in their twistor indices. We define the antisymmetric product of gamma matrices as:
\begin{align}
 \gamma^{IJ}_{ab} \equiv \gamma^{[I}_{ac}\gamma^{J]c}{}_b \ ,
\end{align}
which is symmetric in the twistor indices $ab$. We can use $\gamma_I^{ab}$ to convert between 4+1d vectors $u^I$ and traceless bitwistors $u^{ab}$ as:
\begin{align}
 u^{ab} = \gamma_I^{ab}u^I \quad ; \quad u^I = -\frac{1}{4}\gamma^I_{ab}u^{ab} \ . \label{eq:conversion_5d}
\end{align}
Similarly, we can use $\gamma_{IJ}^{ab}$ to convert between bivectors $M^{IJ}$ and symmetric twistor matrices $M^{ab}$:
\begin{align}
 M^{ab} = \frac{1}{2}\gamma_{IJ}^{ab}M^{IJ} \quad ; \quad M^{IJ} = \frac{1}{4}\gamma^{IJ}_{ab} M^{ab} \ .
\end{align}

Let us now fix a point $v\in dS_4$ in de Sitter space. Our twistor space, i.e. the Dirac representation of $SO(4,1)$, can be identified with the Dirac representation of the Lorentz group $SO(3,1)$ at $v$. It then decomposes into left-handed and right-handed Weyl representations. The decomposition is accomplished by the pair of projectors:
\begin{align}
 \begin{split}
   P^a{}_b(v) &= \frac{1}{2}\left(\delta^a_b - iv^I\gamma_I{}^a{}_b \right) = \frac{1}{2}\left(\delta^a_b - iv^a{}_b \right) \ ; \\
   \bar P^a{}_b(v) &= \frac{1}{2}\left(\delta^a_b + iv^I\gamma_I{}^a{}_b \right) = \frac{1}{2}\left(\delta^a_b + iv^a{}_b \right) \ . \label{eq:projectors}
 \end{split}
\end{align}
These serve as a $v$-dependent version of the familiar chiral projectors in $\bbR^{1,3}$. As in our treatment of vectors, one can use the $(a,b,\dots)$ indices for both $SO(4,1)$ and $SO(3,1)$ Dirac spinors. In addition, at a fixed point $v\in dS_4$, we will use left-handed $(\alpha,\beta,\dots)$ and right-handed $(\dot\alpha,\dot\beta,\dots)$ Weyl spinor indices, which are taken to imply $P(v)$ and $\bar P(v)$ projections, respectively. Thus, a twistor $Z^a$ at a point $v$ decomposes into Weyl spinors $z^\alpha$ and $\bar z^{\dot\alpha}$. The matrices $P_{ab}(v)$ and $\bar P_{ab}(v)$ serve as the metrics $\epsilon_{\alpha\beta}$ and $\epsilon_{\dot\alpha\dot\beta}$ for the respective Weyl spinor spaces.

For a vector $\xi^I$ in the 3+1d tangent space at a de Sitter point $v$, the non-vanishing chiral components of the bitwistor $\xi^{ab}$ are $\xi^{\alpha\dot\beta} = -\xi^{\dot\beta\alpha}$. The conversion formula \eqref{eq:conversion_5d} then becomes:
\begin{align}
 \xi^{\alpha\dot\alpha} = \gamma_I^{\alpha\dot\alpha}\xi^I \quad ; \quad \xi^I = -\frac{1}{2}\gamma^I_{\alpha\dot\alpha}\xi^{\alpha\dot\alpha} \ . \label{eq:conversion_4d}
\end{align}
Similarly, for a bivector $M^{IJ}$ in the tangent space at $v$, the symmetric twistor $M^{ab}$ decomposes into left-handed and right-handed pieces $m^{\alpha\beta},\bar m^{\dot\alpha\dot\beta}$:
\begin{align}
 m^{\alpha\beta} = \frac{1}{2}\gamma_{IJ}^{\alpha\beta}M^{IJ} \ ; \quad \bar m^{\dot\alpha\dot\beta} = \frac{1}{2}\gamma_{IJ}^{\dot\alpha\dot\beta}M^{IJ} \ ; \quad 
 M^{IJ} = \frac{1}{4}\left(\gamma^{IJ}_{\alpha\beta}\, m^{\alpha\beta} + \gamma^{IJ}_{\dot\alpha\dot\beta}\, \bar m^{\dot\alpha\dot\beta} \right) \ . 
\end{align}
Finally, for a bivector $M^{IJ} = 2v^{[I}\xi^{J]}$ with one radial and one tangential index, the symmetric twistor $M^{ab}$ has non-vanishing components $m^{\alpha\dot\beta} = m^{\dot\beta\alpha} = i\xi^{\alpha\dot\beta}$, where $\xi^{\alpha\dot\beta} = -\xi^{\dot\beta\alpha}$ are the spinor components of the tangent vector $\xi^I$:
\begin{align}
 \gamma_{IJ}^{\alpha\dot\beta}\,v^I\xi^J = i\gamma_I^{\alpha\dot\beta}\xi^I \ .
\end{align}
Further details and identities may be found in \cite{Neiman:2013hca,Neiman:2014npa}. 

\section{Linearized higher-spin theory} \label{sec:free}

As a build-up towards our main result, we will now present various formulations of \emph{linearized} higher-spin theory on a de Sitter background. In section \ref{sec:free:constant}, we present the standard formulation with a spacetime-independent compensator. In section \ref{sec:free:non_constant}, we make the compensator spacetime-dependent, specializing the approach in \cite{Didenko:2012vh} to the spinor variables of the 3+1d theory. In the process, the higher-spin gauge invariance is obscured. We restore manifest gauge invariance in section \ref{sec:free:covariant}. This involves introducing much of the machinery of the non-linear theory, and forms a key step towards the non-linear equations in section \ref{sec:interacting}.

\subsection{Constant compensator} \label{sec:free:constant}

The higher-spin algebra for the free theory consists of functions $f(Y)$ of twistor variables $Y^a$, subject to the star product:
\begin{align}
 f\star g = f \exp\left(iI^{ab}\overleftarrow{\frac{\del}{\del Y^a}}\overrightarrow{\frac{\del}{\del Y^b}}\right) g \ . \label{eq:star_Y}
\end{align}
This product is associative, but non-commutative and non-local in $Y$ space. The algebra \eqref{eq:star_Y} contains the generators of the de Sitter group $SO(4,1)$, with the appropriate commutation relations:
\begin{align}
 T_{IJ} = \frac{i}{8}\gamma_{IJ}^{ab}\,Y_a Y_b \quad ; \quad [T^{IJ}, T_{KL}]_\star = 4\delta^{[I}_{[K}\, T^{J]}{}_{L]} \ . \label{eq:generators}
\end{align} 

We view the flat 4+1d indices $(I,J,\dots)$ and their spinor/twistor counterparts $(a,b,\dots)$ as living in an internal $\bbR^{1,4}$ at each point in spacetime. For spacetime points themselves, we will use 4d coordinates $x^\mu$. Fields in spacetime that depend also on the internal twistor variable $Y$ are known as ``master fields''. 

Later on, we will have the option of identifying spacetime with the $dS_4$ hyperboloid \eqref{eq:dS} in the internal space. For now, we instead fix our ``position'' in the internal space by choosing a spacetime-independent vector $v^I$ on the hyperboloid \eqref{eq:dS}. This will be the ``compensator'' responsible for breaking $SO(4,1)$ down to the Lorentz group $SO(3,1)$. 

Thus, for now, the empty de Sitter background for the linearized theory is distinct from the natural $dS_4$ in the internal space. Instead, the structure of the background is given by a connection master field $\Omega(x;Y) = dx^\mu \Omega_\mu(x;Y)$ in spacetime. This background connection satisfies the flatness condition:
\begin{align}
 d\Omega + \Omega\star\Omega = 0 \ , \label{eq:flat_Omega}
\end{align}
where products of forms are always understood to involve a wedge product. We constrain the master field $\Omega(x;Y)$ to be even under $Y^a\rightarrow -Y^a$, so that it contains only integer-spin coefficients in a Taylor expansion in $Y^a$. To interpret these coefficients, we must decompose the twistor $Y^a$ into Weyl spinors $y^\alpha,\bar y^{\dot\alpha}$. This is accomplished by the projectors \eqref{eq:projectors} at our preferred point $v^I$ in the internal space. We then identify the coefficient of $y\bar y$ in $\Omega(x;Y) = \Omega(x;y,\bar y)$ as the spacetime vielbein, while the coefficients of $yy$ and $\bar y\bar y$ encode the spin connection. More generally, the coefficient of $(y)^{s-1}(\bar y)^{s-1}$ encodes the spin-$s$ gauge potential, while all other coefficients are related by \eqref{eq:flat_Omega} to spacetime derivatives of these potentials. Eq. \eqref{eq:flat_Omega} also ensures that these component fields correspond to an empty de Sitter background, up to gauge transformations.

The linearized field strengths of the dynamical higher-spin gauge fields, along with the dynamical spin-zero field, are contained in the scalar master field $B(x;Y)$. As with $\Omega$, we constrain $B$ to be even under $Y^a\rightarrow -Y^a$. The linearized field equations read:
\begin{align}
 dB + \Omega\star B - B\star\tilde\Omega = 0 \ , \label{eq:free_B}
\end{align}
where the ``$\sim$'' operation is defined as:
\begin{align}
 \tilde f(y,\bar y) = f(-y,\bar y) \ , \label{eq:tilde_y}
\end{align}
or, making the $v$-dependence explicit:
\begin{align}
 \tilde f(Y^a) = f(iv^a{}_b Y^b) \ . \label{eq:tilde_Y}
\end{align}

The Taylor expansion of $B(x;y,\bar y)$ in the spinor variables can again be interpreted in terms of component fields. The coefficients of $(y)^{2s}$ and $(\bar y)^{2s}$ encode the left-handed and right-handed pieces of the linearized spin-$s$ field strength, or the scalar field for $s=0$. The field equations for all these field strengths are encoded in \eqref{eq:free_B}. The other Taylor coefficients in $B(x;y,\bar y)$ are related through \eqref{eq:free_B} to spacetime derivatives of the field strengths. 

Eqs. \eqref{eq:flat_Omega}-\eqref{eq:free_B} are invariant under the higher-spin gauge transformations with gauge parameter $\varepsilon(x;Y)$ (again, even under $Y^a\rightarrow -Y^a$):
\begin{align}
 \delta\Omega &= \varepsilon\star\Omega - \Omega\star\varepsilon - d\varepsilon \ ; \label{eq:delta_Omega} \\
 \delta B &= \varepsilon\star B - B\star\tilde\varepsilon \ . \label{eq:delta_B}
\end{align}
Due to the ``$\sim$'' operation in eqs. \eqref{eq:free_B} and \eqref{eq:delta_B}, the master field $B$ is said to live in the ``twisted-adjoint'' representation of the higher-spin algebra.

\subsection{Spacetime-dependent compensator} \label{sec:free:non_constant}

The above formulation of the free theory can be extended to allow for a spacetime-dependent compensator $v^I=v^I(x)$, with gradient $dv^I$. Since $v^I$ must remain a unit vector, we have $v_I dv^I = 0$, i.e. $dv^I$ is in the tangent space to the internal $dS_4$ at $v^I$.

Following \cite{Didenko:2012vh}, we deform the field equation \eqref{eq:free_B} into:
\begin{align}
 dB + \Omega\star B - B\star\tilde\Omega = -2B\star T_{IJ}\,v^I dv^J \ . \label{eq:free_B_T}
\end{align}
The flatness condition \eqref{eq:flat_Omega} stays unchanged. Using our expression \eqref{eq:generators} for the $SO(4,1)$ generators $T_{IJ}$, the field equation \eqref{eq:free_B_T} becomes:
\begin{align}
 dB + \Omega\star B - B\star\tilde\Omega = -\frac{i}{4}\,B\star \left(v^a{}_c\, dv^{cb}\, Y_a Y_b\right) \ , \label{eq:free_B_Y}
\end{align}
where we converted the vector indices on $v^I$ and $dv^I$ into twistor indices via \eqref{eq:conversion_5d}. Converting further into Weyl-spinor indices using the $v$-dependent projectors \eqref{eq:projectors}, this becomes:
\begin{align}
 dB + \Omega\star B - B\star\tilde\Omega = \frac{1}{2}\,B\star \left(dv^{\alpha\dot\alpha} y_\alpha \bar y_{\dot\alpha}\right) \ . \label{eq:free_B_y}
\end{align}
Note that in any case, these is an additional $v$-dependence in the ``$\sim$'' operation. 

As elaborated in \cite{Vasiliev:2001wa,Didenko:2012vh}, eq. \eqref{eq:free_B_T} still encodes free massless equations for field strengths on a de Sitter background. However, the de Sitter vielbein and spin connection are no longer given simply by Taylor coefficients of $\Omega(x;Y)$, but by a combination of these with the compensator gradient $dv^I$. Thus, we can choose both different gauges for $\Omega(x;Y)$ and different compensator functions $v^I(x)$, with each choice leading to a different realization of the de Sitter background. The simplest choice is to set $\Omega=0$ and to \emph{identify} the spacetime coordinates $x^\mu$ with the value of the compensator $v^I$. Our spacetime thus becomes identified with the internal de Sitter space \eqref{eq:dS}, inheriting its geometry. The field equation \eqref{eq:free_B_y} becomes simply:
\begin{align}
 \del_{\alpha\dot\alpha}B = -B\star (y_\alpha \bar y_{\dot\alpha}) \ .
\end{align}

\subsection{Restoring manifest gauge invariance} \label{sec:free:covariant}

The field equation \eqref{eq:free_B_Y} with a spacetime-dependent compensator is no longer manifestly invariant under the gauge transformations \eqref{eq:delta_Omega}-\eqref{eq:delta_B}. This is due to the appearance of an explicit function of $Y^a$ on the RHS, as well as the implicit dependence of the ``$\sim$'' operation on $v^I(x)$. We will now restore manifest gauge invariance by introducing a second twistor variable $Z^a$, with Weyl-spinor components $z^\alpha,\bar z^{\dot\alpha}$. We define the star product for functions $f(Y,Z)$ as:
\begin{align}
 f\star g = f \exp\left(iI^{ab}\left(\overleftarrow{\frac{\del}{\del Y^a}} + \overleftarrow{\frac{\del}{\del Z^a}} \right)
    \left(\overrightarrow{\frac{\del}{\del Y^b}} - \overrightarrow{\frac{\del}{\del Z^b}} \right)\right) g \ . \label{eq:star_YZ} 
\end{align}
This is of course the standard star product from the non-linear Vasiliev theory. It is associative and reduces to the product \eqref{eq:star_Y} for $Z$-independent functions. Crucially, $Y$ and $Z$ commute under the product \eqref{eq:star_YZ}.

We now define the Klein operators:
\begin{align}
 \kappa \equiv \exp\left(iP_{ab}(v)Z^a Y^b\right) = \exp(iy_\alpha z^\alpha) \quad ; \quad 
 \bar\kappa \equiv \exp\left(i\bar P_{ab}(v)Z^a Y^b\right) = \exp(i\bar y_{\dot\alpha} \bar z^{\dot\alpha}) \ . \label{eq:kappa}
\end{align}
Note that $\kappa$ and $\bar\kappa$ depend on the compensator $v^I$, but their product does not:
\begin{align}
 \kappa\bar\kappa = \kappa\star\bar\kappa = \bar\kappa\star\kappa = \exp(iY_a Z^a) \ .
\end{align}
We also have:
\begin{align}
 \kappa\star\kappa = \bar\kappa\star\bar\kappa = 1 \ .
\end{align}

The ``$\sim$'' operation from \eqref{eq:tilde_y} can be extended to functions $f(Y,Z)$ as:
\begin{align}
 \tilde f(y,\bar y,z,\bar z) = f(-y,\bar y,-z,\bar z) \ , \quad \text{i.e.} \quad \tilde f(Y,Z) = f\left(iv^a{}_b Y^a, iv^a{}_b Z^b\right) \ .
\end{align}
Using the Klein operators \eqref{eq:kappa}, this operation can be expressed in terms of star products:
\begin{align}
 \kappa\star f(Y,Z)\star\kappa = \tilde f(Y,Z) \quad ; \quad \bar\kappa\star f(Y,Z)\star\bar\kappa = \tilde f(-Y,-Z) \ . 
\end{align}
As a corollary, functions even under $(Y,Z)\rightarrow(-Y,-Z)$ commute with $\kappa\bar\kappa$, while odd functions anticommute with it. We also see that for a master field $B$ transforming in the twisted-adjoint representation \eqref{eq:delta_B} of the higher-spin algebra, the products $B\star\kappa$ and $B\star\bar\kappa$ transform in the adjoint:
\begin{align}
 \delta(B\star\kappa) = \left[\varepsilon, B\star\kappa\right]_\star \quad ; \quad \delta(B\star\bar\kappa) = \left[\varepsilon, B\star\bar\kappa\right]_\star . \label{eq:delta_B_kappa}
\end{align}

We are now ready to rewrite the linearized field equations with $v^I=v^I(x)$ in a manifestly gauge-covariant form. We continue to keep the master fields $\Omega(x;Y)$ and $B(x;Y)$ independent of $Z$. The flatness condition \eqref{eq:flat_Omega} stays unchanged. We rewrite the field equation \eqref{eq:free_B_Y} as:
\begin{align}
 d(B\star\kappa) + \left[\Omega, B\star\kappa\right]_\star = -\frac{i}{4}\left(v^a{}_c\, dv^{cb} Z_a Z_b\right)\star B\star\kappa \ , \label{eq:free_B_Z}
\end{align}
or, equivalently, with $\bar\kappa$ in place of $\kappa$. The difference in the RHS between \eqref{eq:free_B_Y} and \eqref{eq:free_B_Z} is due to the gradient $d\kappa$, which stems from the dependence of $\kappa$ on the compensator $v^I$. Eq. \eqref{eq:free_B_Z} is manifestly invariant under the transformations \eqref{eq:delta_Omega}-\eqref{eq:delta_B} with a $Z$-independent gauge parameter $\varepsilon(x;Y)$.

Finally, since $\kappa$ anticommutes with $v^a{}_c\, dv^{cb}\, Z_a Z_b \sim dv^{\alpha\dot\alpha}z_\alpha\bar z_{\dot\alpha}$, we can absorb the RHS of \eqref{eq:free_B_Z} into a redefinition of the connection:
\begin{align}
 W \equiv \Omega + \frac{i}{8}\,v^a{}_c\, dv^{cb} Z_a Z_b = \Omega - \frac{1}{4}\,dv^{\alpha\dot\alpha} z_\alpha \bar z_{\dot\alpha} \ . \label{eq:W}
\end{align}
The flatness condition \eqref{eq:flat_Omega} and the field equation \eqref{eq:free_B_Z} become:
\begin{gather}
 dW + W\star W = \frac{i}{16}\,dv^a{}_c\,dv^{cb}Z_a Z_b \ ; \label{eq:flat_W} \\
 d(B\star\kappa) + \left[W, B\star\kappa\right]_\star = 0 \ \label{eq:free_B_W} .
\end{gather}
Thus, the cost of absorbing the RHS of \eqref{eq:free_B_Z} is that the new connection $W$ has a nonzero curvature. Working with $W$ instead of $\Omega$ turns out to simplify the non-linear equations below.

\section{The non-linear theory} \label{sec:interacting}

In this section, we construct the non-linear theory with $v^I = v^I(x)$. First, we allow the master fields $W,B$ to depend arbitrarily on the extra twistor variable $Z$ (prior to imposing the field equations). In addition, we introduce an auxiliary twistor-valued master field $S_a$. Thus, the full set of master fields is:
\begin{align}
 W = dx^\mu W_\mu(x;Y,Z) \ ; \quad B = B(x;Y,Z) \ ; \quad S_a = S_a(x;Y,Z) \ .
\end{align}
We again restrict to the integer-spin sector by making $W,B$ even and $S_a$ odd under $(Y,Z)\rightarrow(-Y,-Z)$. The master fields are subject to gauge transformations with a gauge parameter $\varepsilon(x;Y,Z)$, restricted to be even under $(Y,Z)\rightarrow(-Y,-Z)$:
\begin{align}
 \begin{split}
   \delta W &= \varepsilon\star W - W\star\varepsilon - d\varepsilon \ ; \\
   \delta B &= \varepsilon\star B - B\star\kappa\star\varepsilon\star\kappa = \varepsilon\star B - B\star\bar\kappa\star\varepsilon\star\bar\kappa \ ; \\
   \delta S_a &= \varepsilon\star S_a - S_a\star\varepsilon \ .
 \end{split} \label{eq:gauge}
\end{align}
Thus, $W$ transforms as a connection, $S_a$ transforms in the adjoint, and $B$ transforms in the twisted-adjoint. 

In the linearized limit, we fix $S_a$ to the background value $Z_a$. Thus, the linearized limit is defined by:
\begin{align}
 \begin{split}
   W(x;Y,Z) \ &\longrightarrow \ \Omega(x;Y) + \frac{i}{8}\,v^a{}_c\, dv^{cb} Z_a Z_b \ ; \\
   B(x;Y,Z) \ &\longrightarrow \ \text{small} \ B(x;Y) \ ; \\ 
   S_a(x;Y,Z) \ &\longrightarrow \ Z_a \ .
 \end{split} \label{eq:linearized_limit}
\end{align}
Our task now is to find non-linear field equations with the following properties:
\begin{enumerate}
 \item They should be invariant under spacetime diffeomorphisms and the higher-spin gauge transformations \eqref{eq:gauge}.
 \item In the linearized limit \eqref{eq:linearized_limit}, they should reduce to eqs. \eqref{eq:flat_W}-\eqref{eq:free_B_W}.
 \item In the limit $dv^I = 0$, they should reduce to the standard form of Vasiliev's equations.
 \item The terms proportional to $dv^I$ should not spoil the consistency of Vasiliev's equations. This means that applying an exterior derivative to the equations shouldn't generate any additional constraints.
\end{enumerate}
These properties are all satisfied by the following system:
\begin{gather}
 dW + W\star W = \frac{i}{16}\,dv^a{}_c\,dv^{cb}\,S_a\star S_b \ ; \label{eq:Phi_S} \\
 d(B\star\kappa) + \left[W, B\star\kappa\right]_\star = 0 \ ; \label{eq:D_B_S} \\
 dS_a + \left[W, S_a\right]_\star = -\frac{1}{2}\,v^b{}_c\, dv^c{}_a\,S_b \ ; \label{eq:D_S} \\
 S_a\star B\star\kappa - iv^b{}_a B\star\kappa\star S_b = 0 \ ; \label{eq:S_B} \\
 \left[S_a, S_b\right]_\star = -i\left(2I_{ab} + (I_{ab} - iv_{ab})F_\star(B\star\kappa) + (I_{ab} + iv_{ab})\bar F_\star(B\star\bar\kappa) \right) \ , \label{eq:SS}
\end{gather}
where we recall that $\kappa,\bar\kappa$ depend implicitly on $v^I(x)$. There is some freedom in fixing the interactions, which is encoded in the odd function $F(u)$, with complex conjugate $\bar F(u)$. The subscript in $F_\star(B\star\kappa)$ and $\bar F_\star(B\star\bar\kappa)$ means that the products in the function's Taylor expansion should be interpreted as star-products. There is some redundancy in the choice of $F(u)$ due to field redefinitions, but this will not concern us here. The theory is parity-invariant only for $F(u)=u$ or $F(u) = iu$, with a parity-even or parity-odd scalar field, respectively. Note that the way in which $S_a$ appears on the RHS of \eqref{eq:Phi_S},\eqref{eq:D_S} precludes its standard interpretation as a connection $dZ^a S_a$ in $Z$ space. 

The consistency of eqs. \eqref{eq:Phi_S}-\eqref{eq:SS} will be proven in the Appendix. For the remainder of this section, we will describe some of their properties. Eq. \eqref{eq:Phi_S} has the form of the background equation \eqref{eq:flat_W} from the linearized theory, with $Z_a Z_b$ replaced by $S_{(a}\star S_{b)}$ (the symmetrization arising from the $dv^a{}_c\,dv^{cb}$ factor). Eq. \eqref{eq:D_B_S} is identical to the linearized equation \eqref{eq:free_B_Z}. Eqs. \eqref{eq:D_S}-\eqref{eq:SS} become identities in the linearized limit. The differences from the standard $v^I=\const$ formulation lie in the RHS of eqs. \eqref{eq:Phi_S},\eqref{eq:D_S}. In particular, the constraints \eqref{eq:S_B}-\eqref{eq:SS} on the master fields at a single spacetime point are the same as in the standard formulation. 

As usual, eq. \eqref{eq:SS} expresses $B$ in terms of $S_a$, so that eqs. \eqref{eq:D_B_S},\eqref{eq:S_B} can be viewed as Bianchi identities rather than independent equations. In addition, we will prove in the Appendix that for non-degenerate $dv^I$, the mixed-handedness components of \eqref{eq:SS} are also not independent, but arise from the consistency conditions of eqs. \eqref{eq:Phi_S},\eqref{eq:D_S}. This makes our formulation more economical than the standard one with $v^I = \const$. 

Let us now rewrite the field equations \eqref{eq:Phi_S}-\eqref{eq:SS} in Weyl-spinor notation. First, we decompose $S_a$ into left-handed and right-handed components:
\begin{align}
 s_\alpha \equiv P^b{}_\alpha(v) S_b \quad ; \quad \bar s_{\dot\alpha} \equiv \bar P^b{}_{\dot\alpha}(v) S_b \ .
\end{align}
To describe spacetime derivatives of these chiral components, we define a ``covariant derivative'' operator $\nabla$:
\begin{align}
 \nabla s_\alpha \equiv P^b{}_\alpha(v)\, d(P^c{}_b(v) S_c) \quad ; \quad \nabla\bar s_{\dot\alpha} \equiv \bar P^b{}_{\dot\alpha}(v)\, d(\bar P^c{}_b(v) S_c) \ . \label{eq:nabla}
\end{align}
The field equations \eqref{eq:Phi_S}-\eqref{eq:SS} now become:
\begin{gather}
 dW + W\star W = -\frac{i}{16}\left(dv^\alpha{}_{\dot\gamma}\, dv^{\beta\dot\gamma} s_\alpha\star s_\beta
   + dv_\gamma{}^{\dot\alpha}\, dv^{\gamma\dot\beta}\, \bar s_{\dot\alpha}\star\bar s_{\dot\beta} \right) \ ; \label{eq:Phi_s} \\
 d(B\star\kappa) + \left[W, B\star\kappa\right]_\star = 0 \ ; \label{eq:dB_s} \\
 \nabla s_\alpha + \left[W, s_\alpha\right]_\star = 0 \ ; \quad
 \nabla \bar s_{\dot\alpha} + \left[W, \bar s_{\dot\alpha}\right]_\star = 0 \ ; \label{eq:nabla_s} \\
 \left[s_\alpha, B\star\bar\kappa\right]_\star = 0 \ ; \quad \left[\bar s_{\dot\alpha}, B\star\kappa\right]_\star = 0 \ ; \\
 s_\alpha\star s^\alpha = 2i\big(1 + F_\star(B\star\kappa)\big) \ ; \quad \bar s_{\dot\alpha}\star \bar s^{\dot\alpha} = 2i\big(1 + \bar F_\star(B\star\bar\kappa)\big) \ ;
 \quad \left[s_\alpha, \bar s_{\dot\alpha}\right]_\star = 0 \ . \label{eq:ss}
\end{gather}
The difference in the RHS between eqs. \eqref{eq:D_S} and \eqref{eq:nabla_s} is due to the derivatives acting on the projectors in \eqref{eq:nabla}. As stated above, for non-degenerate $dv^I$, the last equation in \eqref{eq:ss} arises from the consistency conditions of \eqref{eq:Phi_s} and \eqref{eq:nabla_s}, and in that sense is not an independent equation.  Thus, the minimal set of equations from which all others follow reads:
\begin{gather}
 dW + W\star W = -\frac{i}{16}\left(dv^\alpha{}_{\dot\gamma}\, dv^{\beta\dot\gamma} s_\alpha\star s_\beta
   + dv_\gamma{}^{\dot\alpha}\, dv^{\gamma\dot\beta}\, \bar s_{\dot\alpha}\star\bar s_{\dot\beta} \right) \ ; \label{eq:minimal_Phi} \\
 \nabla s_\alpha + \left[W, s_\alpha\right]_\star = 0 \ ; \quad
 \nabla \bar s_{\dot\alpha} + \left[W, \bar s_{\dot\alpha}\right]_\star = 0 \ ; \\
 F^{-1}_\star\!\left(\frac{s_\alpha\star s^\alpha}{2i} - 1 \right)\star\kappa 
   = \bar F^{-1}_\star\!\left(\frac{\bar s_{\dot\alpha}\star\bar s^{\dot\alpha}}{2i} - 1 \right)\star\bar\kappa \ , \label{eq:minimal_ss}
\end{gather}
where we solved for $B$ using the inverse functions of $F(u)$ and $\bar F(u)$.

As in the linearized case, we are now at liberty to identify the spacetime coordinates $x^\mu$ with the values of the compensator $v^I$ in the internal de Sitter space \eqref{eq:dS}. The field equations \eqref{eq:Phi_s}-\eqref{eq:ss} then become:
\begin{gather}
 \nabla^{(\alpha}{}_{\dot\gamma}W^{\beta)\dot\gamma} + W^{(\alpha}{}_{\dot\gamma}\star W^{\beta)\dot\gamma} = -\frac{i}{2} s^{(\alpha}\star s^{\beta)} \ ; \label{eq:phi_on_dS} \\
 \nabla_{\gamma(\dot\alpha}W^\gamma{}_{\dot\beta)} + W_{\gamma(\dot\alpha}\star W^\gamma{}_{\dot\beta)} = -\frac{i}{2} \bar s_{(\dot\alpha}\star\bar s_{\dot\beta)} \ ; \nonumber \\   
 \nabla_{\alpha\dot\alpha}(B\star\kappa) + \left[W_{\alpha\dot\alpha}, B\star\kappa\right]_\star = 0 \ ; \label{eq:D_B_on_dS} \\
 \nabla_{\alpha\dot\alpha}s_\beta + \left[W_{\alpha\dot\alpha}, s_\beta\right]_\star = 0 \ ; \quad 
 \nabla_{\alpha\dot\alpha}\bar s_{\dot\beta} + \left[W_{\alpha\dot\alpha}, \bar s_{\dot\beta}\right]_\star = 0 \ ; \\
 \left[s_\alpha, B\star\bar\kappa\right]_\star = 0 \ ; \quad \left[\bar s_{\dot\alpha}, B\star\kappa\right]_\star = 0 \ ; \\
 s_\alpha\star s^\alpha = 2i\big(1 + F_\star(B\star\kappa)\big) \ ; \quad \bar s_{\dot\alpha}\star\bar s^{\dot\alpha} = 2i\big(1 + \bar F_\star(B\star\bar\kappa)\big) \ ;
 \quad \left[s_\alpha, \bar s_{\dot\alpha}\right]_\star = 0 \ . \label{eq:ss_on_dS}
\end{gather}
Here, $W_{\alpha\dot\alpha}$ are the components of the higher-spin connection $W = -\frac{1}{2}dx^{\alpha\dot\alpha}W_{\alpha\dot\alpha}$, while $\nabla_{\alpha\dot\alpha}$ is the covariant derivative for scalars, spinors and vectors in $dS_4$. When acting on $s_\alpha$ and $\bar s_{\dot\alpha}$, this coincides with the derivative defined in \eqref{eq:nabla}, i.e. $\nabla = -\frac{1}{2}dx^{\alpha\dot\alpha}\nabla_{\alpha\dot\alpha}$.

\subsection{Comparison with the standard Vasiliev equations}

The field equations of standard Vasiliev theory with $v^I=\const$ are obtained by setting $dv^I = 0$ in eqs. \eqref{eq:Phi_s}-\eqref{eq:ss}:
\begin{gather}
dW + W\star W = 0 \ ; \label{eq:standard_Phi} \\
d(B\star\kappa) + \left[W, B\star\kappa\right]_\star = 0 \ ; \\
ds_\alpha + \left[W, s_\alpha\right]_\star = 0 \ ; \quad
d\bar s_{\dot\alpha} + \left[W, \bar s_{\dot\alpha}\right]_\star = 0 \ ; \\
\left[s_\alpha, B\star\bar\kappa\right]_\star = 0 \ ; \quad \left[\bar s_{\dot\alpha}, B\star\kappa\right]_\star = 0 \ ; \\
s_\alpha\star s^\alpha = 2i\big(1 + F_\star(B\star\kappa)\big) \ ; \quad \bar s_{\dot\alpha}\star \bar s^{\dot\alpha} = 2i\big(1 + \bar F_\star(B\star\bar\kappa)\big) \ ;
\quad \left[s_\alpha, \bar s_{\dot\alpha}\right]_\star = 0 \ , \label{eq:standard_ss}
\end{gather}
where we note that for $v^I=\const$, the derivatives \eqref{eq:nabla} of the chiral components $s_\alpha,\bar s_{\dot\alpha}$ reduce to ordinary derivatives $ds_\alpha,d\bar s_{\dot\alpha}$. The standard equations \eqref{eq:standard_Phi}-\eqref{eq:standard_ss} differ from the modified ones \eqref{eq:Phi_s}-\eqref{eq:ss} already at leading order in the field strength. In particular, the RHS of eq. \eqref{eq:Phi_s} is non-vanishing already at zeroth order, while in \eqref{eq:standard_Phi} it is absent. The reason for this difference is the different encoding of the zeroth-order pure-$dS_4$ geometry, in particular the shift \eqref{eq:W} of the higher-spin connection. By reversing this shift, one can remove the leading-order difference between the two formulations, at the cost of complicating the full non-linear equations. Specifically, replacing $Z_a \rightarrow S_a$ in \eqref{eq:W} as required for the non-linear theory, we can define the connection:
\begin{align}
 W' \equiv W + \frac{1}{4}dv^{\alpha\dot\alpha}s_\alpha\star\bar s_{\dot\alpha} = \Omega + O(B) \ .
\end{align}
In terms of $W'$, our field equations \eqref{eq:Phi_s}-\eqref{eq:ss} become:
\begin{gather}
 dW' + W'\star W' = \frac{i}{16}\left(dv^\alpha{}_{\dot\gamma}\, dv^{\beta\dot\gamma} s_\alpha\star s_\beta\star\bar F_\star(B\star\bar\kappa)
    + dv_\gamma{}^{\dot\alpha}\, dv^{\gamma\dot\beta}\, \bar s_{\dot\alpha}\star\bar s_{\dot\beta}\star F_\star(B\star\kappa) \right) \ ; \label{eq:complicated_Phi} \\
 d(B\star\kappa) + \left[W', B\star\kappa\right]_\star = \frac{1}{2}\,dv^{\alpha\dot\alpha}s_\alpha\star\bar s_{\dot\alpha}\star B\star\kappa \ ; \label{eq:complicated_dB} \\
 \nabla s_\alpha + \left[W', s_\alpha\right]_\star = \frac{i}{2}\,dv_\alpha{}^{\dot\alpha}\, \bar s_{\dot\alpha}\star\big(1 + F_\star(B\star\kappa)\big) \ ; \label{eq:complicated_ds} \\
 \nabla \bar s_{\dot\alpha} + \left[W', \bar s_{\dot\alpha}\right]_\star = \frac{i}{2}\,dv^\alpha{}_{\dot\alpha}\, s_\alpha\star\big(1 + \bar F_\star(B\star\bar\kappa)\big) \ ; \label{eq:complicated_ds_bar} \\
 \left[s_\alpha, B\star\bar\kappa\right]_\star = 0 \ ; \quad \left[\bar s_{\dot\alpha}, B\star\kappa\right]_\star = 0 \ ; \\
 s_\alpha\star s^\alpha = 2i\big(1 + F_\star(B\star\kappa)\big) \ ; \quad \bar s_{\dot\alpha}\star \bar s^{\dot\alpha} = 2i\big(1 + \bar F_\star(B\star\bar\kappa)\big) \ ; 
    \quad \left[s_\alpha, \bar s_{\dot\alpha}\right]_\star = 0 \ . \label{eq:complicated_ss}
\end{gather}
The RHS in eq. \eqref{eq:complicated_Phi} is now manifestly proportional to the field strength. Eq. \eqref{eq:complicated_dB} develops a non-vanishing RHS, analogous to that of the linearized eq. \eqref{eq:free_B_Z}. Eqs. \eqref{eq:complicated_ds}-\eqref{eq:complicated_ds_bar} seemingly develop an RHS which is non-vanishing at zeroth order in the field strength. However, those terms precisely cancel with the difference between covariant and partial derivatives of $s_\alpha,\bar s_{\dot\alpha}$. When eqs. \eqref{eq:complicated_ds}-\eqref{eq:complicated_ds_bar} are written with partial derivatives, their RHS is proportional to the field strength:
\begin{align}
  dS_a + \left[W', S_a\right]_\star = -\frac{i}{4}\, dv^c{}_a\,S_b\left((\delta_c^b + iv^b{}_c)  F_\star(B\star\kappa) - (\delta_c^b - iv^b{}_c) \bar F_\star(B\star\bar\kappa)\right) \ .
\end{align}
This is the closest leading-order agreement we can get with the standard $v^I = \const$ equations. While otherwise cumbersome, the form \eqref{eq:complicated_Phi}-\eqref{eq:complicated_ss} of the equations has one advantage: when written out explicitly to first order in $B$, eq. \eqref{eq:complicated_Phi} makes manifest the relation between the first-order piece of the connection $W'(x;Y,0)$ and the field strength $B(x;Y,0)$.

\section{Discussion} \label{sec:discuss}

In this paper, we extended the field equations of 3+1d higher-spin gravity to allow for a spacetime-dependent compensator $v^I(x)$. This allowed us in particular to identify spacetime with the pure de Sitter space \eqref{eq:dS} of possible $v^I$ values. A key technical detail in our construction is that the twistor-valued  master field $S_a$ can no longer be treated as a connection $dZ^a S_a$ in twistor space. 

A question now arises: have we uncovered a broader class of physically distinct higher-spin theories, or merely a new formulation of Vasiliev theory? Of particular interest are the two limiting cases: the standard formulation \eqref{eq:standard_Phi}-\eqref{eq:standard_ss} with constant $v^I$ and the fixed-background formulation \eqref{eq:phi_on_dS}-\eqref{eq:ss_on_dS}. Both formulations have a linearized limit where they describe free massless fields in de Sitter space. In the $v^I=\const$ formulation, the de Sitter geometry is encoded (in the simplest gauge) in a connection $W(x;Y)=\Omega(x;Y)$ quadratic in $Y^a$. In the fixed-background formulation, it is encoded instead in the structure of the internal space (while $\Omega(x;Y)$ can be gauged to zero, leaving a connection $W(x;Z)$ quadratic in $Z$). The question is then: given these different descriptions of the free-field limit, do the two formulations describe physically distinct interactions? Our conjecture is that the two formulations are in fact physically equivalent. This is based in part on the prejudice that the higher-spin symmetry (with a few extra choices, such as parity properties) should be powerful enough to determine the theory. 

The above conjecture is probably best posed with regard to boundary $n$-point functions on an (A)dS background. In particular, consider $n$-point functions with boundary conditions that preserve the higher-spin symmetry \cite{Vasiliev:2012vf} (this requires a parity-invariant version of the theory, i.e. $F(u) = u$ or $F(u) = iu$). If we \emph{assume} that the $n$-point functions for both formulations are described by a boundary CFT, then they are constrained by the higher-spin symmetry \cite{Maldacena:2011jn} to be those of a \emph{free} CFT, implying that the two formulations must agree. 

It would be interesting to test this conjectured equality of the $n$-point functions. First, one can try and reproduce the 3-point function calculation of \cite{Giombi:2009wh,Giombi:2010vg} in our fixed-background formulation. Second, one can try and reproduce the indirect symmetry-based argument of \cite{Colombo:2012jx,Didenko:2012tv,Didenko:2013bj}, which yields all the $n$-point functions. The latter argument crucially depends on the vanishing covariant derivative of the master field $B$, which allows its spacetime evolution to be expressed as a gauge transformation. The same property serves to simplify the 3-point calculation in \cite{Giombi:2010vg}. We expect that the field equation \eqref{eq:D_B_on_dS} will play a similar role in the fixed-background version of the theory, even though the connection $W$ is no longer flat.

If the higher-spin (A)dS/CFT indeed applies to our fixed-background formulation, the implications are exciting. The main difficulty in dS/CFT is in relating the CFT at infinity to observable physics inside cosmological horizons. This problem is hard in part because gravitational perturbations (as well as gauge transformations in higher-spin theory) can alter the horizon's location and shape. On the other hand, in our new formulation, the spacetime is always pure $dS_4$, so that cosmological horizons are simple once again. This opens a window towards relating dS/CFT to the physics seen by observers.

The application sketched above is only meaningful if the de Sitter horizons in the fixed-background formulation \emph{behave} as horizons in the causal sense. This is an important open question. More generally, are the higher-spin interactions causal? Now that we have a formulation with a gauge-independent spacetime metric, this question can be properly posed.

The extra structure of the background pure de Sitter space may also help in finding an action formalism for higher-spin gravity (for existing attempts, see \cite{Boulanger:2011dd,Boulanger:2012bj}). The fact that the $\left[s_\alpha,\bar s_{\dot\alpha}\right]_\star$ constraint is no longer independent may be useful as well, since there is now one less equation to be generated by the variational principle. Of course, finding an action for the theory would be a step towards quantization, which so far exists only indirectly through the AdS/CFT duality.

Finally, we believe that there's a lesson to be drawn from the way in which the master field $S_a$ appears in our equations. As mentioned above, our generalization of the Vasiliev system is incompatible with interpreting $S_a$ as a connection in $Z$ space. The apparent lesson, implied also by the non-locality of the star product, is that one should avoid thinking locally in $Z$ space. In particular, one should not try to extract the ``physical'' interacting spin-$s$ fields from the master fields at $Z=0$. Instead, we suggest that the notion of individual spin-$s$ fields is a weak-field approximation, valid only when the master fields are well-described (up to gauge transformations) by the linearized limit \eqref{eq:linearized_limit}. Crucially, this approximation \emph{holds} at infinity in a locally asymptotically (A)dS spacetime. This is all that is needed to speak meaningfully of boundary correlation functions in (A)dS/CFT.

\section*{Acknowledgements}

I am grateful to Laurent Freidel, Lucas Hackl, Illan Halpern and Vasudev Shyam for discussions, as well as to Evgeny Skvortsov and Mikhail Vasiliev for email exchanges. Research at Perimeter Institute is supported by the Government of Canada through Industry Canada and by the Province of Ontario through the Ministry of Research \& Innovation. YN also acknowledges support of funding from NSERC Discovery grants. A portion of this work was carried out while visiting UC Berkeley.

\appendix
\section{Consistency analysis of the new non-linear equations}

In this Appendix, we analyze the consistency of our proposed system \eqref{eq:Phi_s}-\eqref{eq:ss}. Our goal is to prove that the equations are consistent, as well as to show that the $\left[s_\alpha, \bar s_{\dot\alpha}\right]_\star$ equation follows from eqs. \eqref{eq:Phi_s},\eqref{eq:nabla_s} when $dv^I$ is non-degenerate. To make the analysis more efficient, we introduce notations for the curvature of $W$ and for the covariant derivative in the adjoint representation:
\begin{align}
 \Phi \equiv dW + W\star W \quad ; \quad D \equiv \nabla + \left[W,\,\right]_\star \ . \label{eq:abbreviations}
\end{align}
For scalars, $\nabla$ in \eqref{eq:abbreviations} is the ordinary exterior derivative $d$. For quantities with spinor indices, we define it as in \eqref{eq:nabla}, i.e. with chiral projections both before and after taking the derivative:
\begin{align}
 \begin{split}
   \nabla f_{\alpha_1\dots\alpha_m\dot\alpha_1\dots\dot\alpha_n} 
     \equiv{}& P^{b_1}{}_{\alpha_1}(v)\dots P^{b_m}{}_{\alpha_m}(v) \bar P^{c_1}{}_{\dot\alpha_1}(v)\dots\bar P^{c_n}{}_{\dot\alpha_n}(v) \\
     &\cdot d\!\left(P^{d_1}{}_{b_1}(v)\dots P^{d_m}{}_{b_m}(v) \bar P^{e_1}{}_{c_1}(v)\dots\bar P^{e_n}{}_{c_n}(v) f_{d_1\dots d_m e_1\dots e_n} \right) \ .
 \end{split}
\end{align}
It is useful to note that the $\nabla$ derivative annihilates $dv^{\alpha\dot\alpha}$:
\begin{align}
 \nabla dv^{\alpha\dot\alpha} = 0 \ . \label{eq:nabla_dv}
\end{align}
Indeed, writing out:
\begin{align}
 \nabla dv^{\alpha\dot\alpha} = P^\alpha{}_b(v) \bar P^{\dot\alpha}{}_c(v)\, d\big(P^b{}_d(v) \bar P^c{}_e(v)\, dv^{de} \big) \ ,
\end{align}
one sees that there are no contributions from derivatives of the chiral projectors, since the same-handedness components $dv^{\alpha\beta}$ and $dv^{\dot\alpha\dot\beta}$ vanish. Note that the $\nabla$ derivative does not square to zero, essentially due to the curvature of the de Sitter space \eqref{eq:dS}. In particular, for spinors such as $s_\alpha$ and $\bar s_{\dot\alpha}$, we have:
\begin{align}
 \nabla^2 s_\alpha = -\frac{1}{4}\,dv_{\alpha\dot\gamma}\,dv^{\beta\dot\gamma} s_\beta \quad ; \quad 
 \nabla^2 \bar s_{\dot\alpha} = -\frac{1}{4}\,dv_{\gamma\dot\alpha}\,dv^{\gamma\dot\beta}\,\bar s_{\dot\beta} \ . \label{eq:nabla_squared}
\end{align}

With the abbreviated notation \eqref{eq:abbreviations}, our field equations \eqref{eq:Phi_s}-\eqref{eq:ss} take the form:
\begin{gather}
 \Phi = -\frac{i}{16}\left(dv^\alpha{}_{\dot\gamma}\, dv^{\beta\dot\gamma} s_\alpha\star s_\beta
   + dv_\gamma{}^{\dot\alpha}\, dv^{\gamma\dot\beta}\, \bar s_{\dot\alpha}\star\bar s_{\dot\beta} \right) \ ; \label{eq:app_Phi} \\
 D(B\star\kappa) = 0 \quad \Longleftrightarrow \quad D(B\star\bar\kappa) = 0 \ ; \label{eq:app_dB} \\
 Ds_\alpha = 0 \ ; \quad D\bar s_{\dot\alpha} = 0 \ ; \label{eq:app_ds} \\
 \left[s_\alpha, B\star\bar\kappa\right]_\star = 0 \ ; \quad \left[\bar s_{\dot\alpha}, B\star\kappa\right]_\star = 0 \quad \Longleftrightarrow \quad
 \left\{s_\alpha, B\star\kappa\right\}_\star = 0 \ ; \quad \left\{\bar s_{\dot\alpha}, B\star\bar\kappa\right\}_\star = 0 \ ; \label{eq:app_sB} \\
 s_\alpha\star s^\alpha = 2i\big(1 + F_\star(B\star\kappa)\big) \ ; \quad \bar s_{\dot\alpha}\star \bar s^{\dot\alpha} = 2i\big(1 + \bar F_\star(B\star\bar\kappa)\big) \ ;
 \quad \left[s_\alpha, \bar s_{\dot\alpha}\right]_\star = 0 \ . \label{eq:app_ss}
\end{gather}
The double-sided arrows in eqs. \eqref{eq:app_dB},\eqref{eq:app_sB} denote equivalent sets of equations. The equivalence is due to $\kappa\bar\kappa$ being $x$-independent and commuting/anticommuting with functions that are even/odd under $(Y,Z)\rightarrow (-Y,-Z)$.

Eqs. \eqref{eq:app_sB}-\eqref{eq:app_ss} are clearly consistent among themselves, since they are the same as in the standard Vasiliev system. It remains to show that no new equations are generated by applying the derivative $D$ to any of \eqref{eq:app_Phi}-\eqref{eq:app_ss}. For \eqref{eq:app_sB}-\eqref{eq:app_ss}, this is clearly the case, since all the ingredients have vanishing $D$ derivatives due to \eqref{eq:app_dB}-\eqref{eq:app_ds}. For eq. \eqref{eq:app_dB}, we need to check the identity:
\begin{align}
 D^2(B\star\kappa) = \left[\Phi, B\star\kappa\right]_\star \ .
\end{align}
The LHS clearly vanishes due to \eqref{eq:app_dB}, while the RHS vanishes due to \eqref{eq:app_Phi},\eqref{eq:app_sB}. 

It remains to see what happens when we apply a $D$ derivative to eqs. \eqref{eq:app_Phi},\eqref{eq:app_ds}. It is instructive to first consider these two equations in isolation, without imposing any of the others. When applying a derivative to \eqref{eq:app_Phi}, we should obtain the identity $D\Phi = 0$. This is indeed the case, due to \eqref{eq:app_ds} and the identity \eqref{eq:nabla_dv}. Finally, when applying a derivative to \eqref{eq:app_ds}, we should get the identities:
\begin{align}
 D^2 s_\alpha = -\frac{1}{4}\,dv_{\alpha\dot\gamma}\,dv^{\beta\dot\gamma} s_\beta + \left[\Phi, s_\alpha\right]_\star \quad ; \quad
 D^2 \bar s_{\dot\alpha} = -\frac{1}{4}\,dv_{\gamma\dot\alpha}\,dv^{\gamma\dot\beta}\,\bar s_{\dot\beta} + \left[\Phi, \bar s_{\dot\alpha}\right]_\star \ , \label{eq:D_squared}
\end{align}
where the first terms are coming from \eqref{eq:nabla_squared}. The LHS of each equation in \eqref{eq:D_squared} is clearly zero due to \eqref{eq:app_ds}. Equating the RHS to zero and substituting \eqref{eq:app_Phi} for $\Phi$, we get:
\begin{align}
 \begin{split}
   0 &= -\frac{i}{16}\left(dv_{\alpha\dot\gamma}\,dv^{\beta\dot\gamma}\left\{s_\beta, s_\gamma\star s^\gamma - 2i \right\}_\star
     - dv_\delta{}^{\dot\beta}\,dv^{\delta\dot\gamma}\left\{\bar s_{\dot\beta}, \left[s_\alpha, \bar s_{\dot\gamma} \right]_\star  \right\}_\star \right) \ ; \\
   0 &= -\frac{i}{16}\left(dv^\beta{}_{\dot\delta}\,dv^{\gamma\dot\delta}\left\{s_\beta, \left[s_\gamma, \bar s_{\dot\alpha} \right]_\star \right\}_\star   
     + dv_{\gamma\dot\alpha}\,dv^{\gamma\dot\beta}\left\{\bar s_{\dot\beta}, \bar s_{\dot\gamma}\star\bar s^{\dot\gamma} - 2i \right\}_\star \right) \ .
 \end{split}
\end{align}
Equating to zero the coefficients of independent $dvdv$ combinations, we get:
\begin{gather}
 \left\{s_\beta, s_\alpha\star s^\alpha - 2i \right\}_\star = 0 \ ; \quad \left\{\bar s_{\dot\beta}, \bar s_{\dot\alpha}\star\bar s^{\dot\alpha} - 2i \right\}_\star = 0 \ ; 
 \label{eq:s_s_s} \\
 \left\{s_\beta, \left[s_\alpha, \bar s_{\dot\alpha} \right]_\star \right\}_\star = 0 \ ; \quad 
 \left\{\bar s_{\dot\beta}, \left[s_\alpha, \bar s_{\dot\alpha} \right]_\star \right\}_\star = 0 \ . \label{eq:s_s_bar_s}
\end{gather}
We now make two observations. First, the equations above are indeed satisfied once we impose eqs. \eqref{eq:app_sB}-\eqref{eq:app_ss}. For the first equation in \eqref{eq:s_s_s}, we notice that according to eq. \eqref{eq:app_ss}: 
\begin{align}
 s_\alpha\star s^\alpha - 2i = 2iF_\star(B\star\kappa) \ .
\end{align}
This indeed anticommutes with $s_\beta$, since the function $F(u)$ is odd, and $B\star\kappa$ anticommutes with $s_\beta$ according to \eqref{eq:app_sB}. The second equation in \eqref{eq:s_s_s} follows analogously. Finally, eq. \eqref{eq:s_s_bar_s} follows from the last equation in \eqref{eq:app_ss}. This concludes our proof that the system \eqref{eq:app_Phi}-\eqref{eq:app_ss} is consistent.

Our second observation regarding eqs. \eqref{eq:s_s_s}-\eqref{eq:s_s_bar_s} is that \emph{their} consistency conditions suffice to derive the equation $\left[s_\alpha, \bar s_{\dot\alpha} \right]_\star = 0$. Indeed, commuting the first equation in \eqref{eq:s_s_s} with $\bar s_{\dot\beta}$ and repeatedly using \eqref{eq:s_s_bar_s}, we get:
\begin{align}
 \begin{split}
   0 &= \left\{\left[s_\beta, \bar s_{\dot\beta} \right]_\star, s_\alpha\star s^\alpha - 2i \right\}_\star 
     + \left\{s_\beta, \left[s_\alpha\star s^\alpha, \bar s_{\dot\beta} \right]_\star \right\}_\star \\
    &= 2\left(s_\alpha\star s^\alpha - 2i \right)\star\left[s_\beta, \bar s_{\dot\beta} \right]_\star 
     + \left\{s_\beta, 2s_\alpha \star \left[s^\alpha, \bar s_{\dot\beta} \right]_\star \right\}_\star \\
    &= 2\left(s_\alpha\star s^\alpha - 2i \right)\star\left[s_\beta, \bar s_{\dot\beta} \right]_\star 
     + 2\left[s_\beta,s_\alpha\right]_\star \star \left[s^\alpha, \bar s_{\dot\beta} \right]_\star \\
    &= 2\left(s_\alpha\star s^\alpha - 2i \right)\star\left[s_\beta, \bar s_{\dot\beta} \right]_\star 
     - 2s_\alpha\star s^\alpha \star \left[s_\beta, \bar s_{\dot\beta} \right]_\star \\
    &= -4i\left[s_\beta, \bar s_{\dot\beta} \right]_\star \ .
 \end{split}
\end{align}
This proves that the entire system \eqref{eq:app_Phi}-\eqref{eq:app_ss} can be reconstructed from \eqref{eq:app_Phi}, \eqref{eq:app_ds} and the first two equations in \eqref{eq:app_ss}. Solving for $B$, we obtain the minimal system \eqref{eq:minimal_Phi}-\eqref{eq:minimal_ss} that was presented in the main text.

\end{document}